\documentclass[preprintnumbers,twocolumn,nofootinbib]{revtex4}
\usepackage{graphicx}
\usepackage{dcolumn}
\usepackage{bm}
\usepackage{epsfig,amsmath}
\usepackage{amssymb}
\usepackage{subfigure}
\usepackage{float}
\usepackage{ulem}
\usepackage[usenames,dvipsnames]{color}
\usepackage[pagebackref=false, colorlinks=true]{hyperref}
\definecolor{redish}{rgb}{0.7,0.2,0.0}  
\definecolor{bluish}{rgb}{0.2,0.5,0.8}

\hypersetup{linkcolor=redish,          
                  citecolor=blue,        
                  filecolor=magenta,      
                  urlcolor=bluish}          

\DeclareFontFamily{U}{rsfs}{}         
\DeclareFontShape{U}{rsfs}{m}{n}{<5> rsfs5 <6><7> rsfs7          %
  <8><9><10><10.95><12><14.4><17.28><20.74><24.88> rsfs10}{}     %
\DeclareMathAlphabet{\mathfs}{U}{rsfs}{m}{n}

\def \f{\frac}

\def \O{\Omega}

\def \p{\partial}

\def \L{\Lambda}

\def \th{\theta}

\newcommand{\be}{\begin{eqnarray}}
\newcommand{\ee}{\end{eqnarray}}

\begin{document}

\title{Gravitational Larmor precession}

\author{Chandrachur Chakraborty}
\email{chandrachur.c@manipal.edu} 
\affiliation{Manipal Centre for Natural Sciences, Manipal Academy of Higher Education, Manipal 576104, India}

\author{Parthasarathi Majumdar}
\email{parthasarathi.majumdar@iacs.res.in}
\affiliation{Indian Association for the Cultivation of Science, Jadavpur, Kolkata 700032, India}

\begin{abstract}
Inspired by the reported existence of substantive magnetic fields in the vicinity of the central supermassive black holes in Sagitarius A* and Messier 87*, we consider test particle motion in the spacetime close to a generic spherical black hole in the presence of  magnetic fields in its vicinity. Modelling such a spacetime in terms of an axisymmetric, non-rotating Ernst-Melvin-Schwarzschild black hole geometry with appropriate parameters, we compute the geodesic nodal-plane precession frequency for a test particle with mass, for such a spacetime, and obtain a non-vanishing result, surpassing earlier folklore that only axisymmetric spacetimes with rotation (non-vanishing Kerr parameter)  can generate such a precession. We call this magnetic field-generated phenomenon Gravitational Larmor Precession. What we present here is a Proof of Concept incipient assay, rather than a detailed analysis of supermassive black holes with magnetic fields in their neighbourhood. However, for completeness, we briefly discuss observational prospects of this precession in terms of available magnetic field strengths close to central black holes in galaxies.
\end{abstract}
\maketitle

\section{Introduction}\label{intro}

Recent observations of unusually large Faraday rotations of radio pulsar emissions as well as synchrotron emission from accretion discs have provided evidence for unexpectedly large magnetic fields in the vicinity of the central supermassive black holes (SMBHs) in galaxies. In particular, for the Milky Way galaxy, emissions from radio pulsar J1745-2900 appear to exhibit unusually large Faraday rotation measures \cite{eatough2013}, ostensibly because of substantive magnetic fields present near the central black hole (BH) Sagitarius A* (Sgr A*). These magnetic fields may be as large as 1 mG, to the extent that astrophysical models for radio pulsar emission near the central BH are accurate. Likewise, the Event Horizon Telescope collaboration, observing the galaxy Messier 87 (M87), have interpreted the observed radiation as largely made up of synchrotron radiation with its linear polarization rendered unresolved due to large Faraday rotation proximitous to the postulated central BH \cite{eht2021}. Inspired by such recent astrophysical observations, the issue of test particle motion in black hole spacetimes in presence of substansive magnetic fields assumes  a good deal of importace. However, a comprehensive astrophysical analysis of the two supermassive black holes mentioned is far beyond our technical abilities. Instead, we would like to focus here in this paper, on a possible novel phenomenon of a test particle undergoing a {\it nodal plane precession} in the vicinity of a {\it non-rotating} generic black hole in presence of magnetic fields in its vicinity. Of course, this requires us to model the spacetime approximately in such a way that the precession frequencies of test particles travelling along geodesics can be easily computed. Thus, ours is a Proof of Concept assay, rather than a truly viable, full astrophysical analysis, and much more work is needed to bridge the gap between the two aspects. Nevertheless, the novelty of our results motivates us to present them to professional colleagues  for their considered opinion.  

An exact class of solutions of Einstein-Maxwell equation with a uniform, constant magnetic field has been derived by Ernst \cite{ernst1976}, building up on a cosmological spacetime constructed first by Melvin. The resulting spacetime is not completely aysmptotically flat. However, in contrast to a generically asymptotically non-flat spacetime, the Ernst-Melvin-Schwarzschild black hole spacetime has some special properties which enable us to choose the frame of observation {\it identically} to that in an asymptotically flat spacetime. As we shall mention in the next section (details in the Appendix \ref{kms} and Appendix \ref{nps}), the Kretschmann scalar and the Newman-Penrose coefficients for the Ricci and Weyl scalars all vanish at radial asymptopia $r \rightarrow \infty$ {\it on the equatorial plane}. The asymptotic vanishing of curvature scalars in a subregion of the EMS spacetime is, as far as we know, never before dicussed in the literature. Thus, the EMS spacetime has an {\it asymptotically flat subspace}. The vanishing of the curvature invariant within a subspace of the EMS spacetime automatically implies that there must exist a coordinate system at asymptotic radial infinity which is Cartesian-like, appropriate to this subspace. The absence of curvature asymptotically within this subspace, in fact, ensures that such a frame must exist. Likewise the bilinear Maxwell invariant will also be seen to vanish in the same plane asymptotically. Since our analysis of geodesic motion is confined to the equatorial plane, we do not anticipate any special need for concern regarding the choice of observers in this case. Therefore, the precession  frequency discussed here  could be measured with respect to a Cartesian-like frame whose axes are aligned with the ``fixed stars'' \cite{ckp, cm, cbgm, cbgm2} at spatial infinity. For instance, one can measure the orbital plane precession frequency by using a `Copernican frame' (see \cite{cm, ckp} for details). Since the Copernican frame does not rotate (by construction) relative to the inertial frames at asymptotic infinity (i.e., the `fixed stars'), the precession rate in the Copernican frame also gives the precession rate of the orbital plane relative to the fixed stars \cite{cm}. Note that one can also measure the precession frequencies in the asymptotically non-flat spacetimes (e.g., magnetized Kerr and magnetized Reissner-Nordstr\"om \cite{cprd1}) by using the `Copernican frame' \cite{cm}. 
 
We further note that the asymptotic non-flatness of the EMS spacetime is only upto the spatial extent of the magnetic field. As explained in \cite{ernst1976}, realistically the magnetic field is expected to be approximately constant only within a definite spatial region around the BH, and decays beyond that region. This region is often known as the `emission region' \cite{eht2021}.  We consider {\it precession} effects on electrically neutral test particles in such a region. This is akin to the Larmor precession of electrically neutral subatomic particles like neutrons in magnetic fields because of their spin magnetic moment \cite{fls}. This justifies the essence of the title of the paper.

In the curved spacetime of general relativity, curvature induces the phenomenon of {\it geodesic} precession, also called de Sitter precession after its discoverer. In addition, for rotating spacetimes like the Kerr BH, for a {\it circular} test geodesic, the orbital plane of the particle precesses due to the Lense-Thirring (LT) effect \cite{lt}. This is known as the nodal plane precession or the orbital plane precession. It was earlier thought that the orbital plane precession arises only in the presence of a non-vanishing Kerr parameter signifying a rotating spacetime. In general, however, it can be easily traced to the stationarity and axial symmetry of the spacetime. For example, the orbital plane precession due to the LT effect arises in the Kerr \cite{ka}, Kerr-Newman \cite{kr21}, Kerr-Newman-de
Sitter \cite{kr21}, Kerr-Newman (anti)de
Sitter \cite{kr14} and other similar spacetimes \cite{ckp, cgm, cm2}. The effect of the external magnetic field on the LT precession of a test gyroscope was also calculated for the Kerr-Newman black hole \cite{riz} and the rotating neutron stars \cite{ccb}. Similarly, one can also deduce the LT precession in the magnetized Kerr spacetime \cite{ew}, and study the effect of the magnetic fields on it. However, here we study the orbital plane precession of non-rotating spacetimes like the Schwarzschild BHs in the presence of {\it magnetic fields}. This is because, the magnetic field automatically provides an axis of symmetry around which stationary BH solutions have been constructed \cite{ernst1976}, with vanishing Kerr parameter. In this paper, we examine electrically neutral test particle circular geodesics in the equatorial plane of the Ernst-Melvin-Schwarzschild (EMS) spacetime. Our focus is on the possible orbital (or nodal) plane precession of such geodesics, generated mainly by the magnetic field with its interplay with curvature. Because of the possible origin of such precession in the EMS geometry from magnetic fields in its vicinity, we call such precession the Gravitational Larmor Precession (GLP). We shall compute the nodal plane precession angular frequencies for equatorial plane circular geodesics in the EMS BH spacetimes. Choosing this to be a model for test particle motion in the vicinity of central BHs Sgr A* and M87*, observational prospects for GLP are on our agenda. In a realistic astrophysical observation, the orbital plane precession frequency could also be measured using the relativistic precession model of quasiperiodic oscillations (QPOs) of X-ray intensity \cite{sv, ckp}.

The rest of the paper is organized as follows : in Sect. \ref{s2}, the EMS spacetime geometry is briefly described, following both the original formulation by Ernst \cite{ernst1976}, and also the clarificatory work of Gal'tsov-Petukov \cite{gp1978}. In the same section, we present formulae for the computation of angular frequencies in terms of metric components. In Sect. \ref{s3}, we present our results, appropriate to SMBHs under discussion. We also provide graphical representations of our results for greater clarity, pointing out the range of validity of the model under use. In this section, we also examine observational prospects for the precession effect considered in this paper, and the manner in which observational accessibility may be possible. In Sect. \ref{dis}, we end with discussions covering both phenomenological and formal aspects of this work, and also our future outlook. As already mentioned, the two Appendices \ref{kms} and \ref{nps} supplement the discussion of the EMS spacetime geometry given in Sect. \ref{s2}, delienating the subspace for which the curvature scalars vanish at asymptopia.      

\section{Essential aspects of the geometry and general relativistic precession \label{s2}}

\subsection{The EMS Spacetime}

The exact electrovac solution (in $G=c=1$ unit) of the Einstein-Maxwell equation for the   Schwarzschild BH with a uniform, constant magnetic field in its vicinity, can be written as  \cite{ernst1976}
\begin{eqnarray}\nonumber
 ds^2 &=& \L^2\left[- \left(1-\f{r_s}{r} \right)dt^2 + \f{dr^2}{1-\f{r_s}{r} }+ r^2 d\th^2 \right]\\
 &+& \L^{-2}r^2\sin^2 \th d\phi^2
 \label{sz}
\end{eqnarray}
where
\begin{eqnarray}
 \L=1 + \f{r^2}{r_B^2} \sin^2\th
 \label{oL}
\end{eqnarray}
and $r_s$ and  $r_B$ are two length parameters of the solution, with $r_s=2M $ being the Schwarzschild radius, and $r_B = 2/B$ with $B$ being the magnetic field parameter characterizing the uniform magnetic field in the emission region\footnote{In the conversion to gravitational units, there is a multiplcative real, dimensionless factor $\lambda \in [0,1]$, so that $r_B = 2 \lambda/B$. $\lambda$ is chosen as per the appropriate astrophysical situation under consideration. Here, $\lambda$ has to be chosen so that $r_B$ approximately provides the extent of the emission region.}  The components of the magnetic field in an orthonormal (Cartan) frame are given by \cite{ernst1976}, \cite{gp1978}
\begin{eqnarray}
 B_r &=& \L^{-2} B \cos\th
 \\
 B_{\th} &=& -\L^{-2} B\left(1-\f{r_s}{r} \right)^{1/2} \sin\th
\end{eqnarray}
where the angular component vanishes upon event horizon: $R_h=2M$. It is obvious that the orthonormal components of magnetic field decay rapidly for $r >> r_B$, so  that the region described by the radial coordinate $r \in [r_s, r_B]$ may be taken to be the {\it emission region} for central SMBHs of galaxies. It is in this region that the magnetic field in the galaxy may be considered to be approximately uniform. In the equatorial plane of spacetime, $\th= \pi/2$, so that $B_{\th}$ is the only component that survives for circular timelike geodesics in this plane. These are the sort of geodesics we shall consider in the next section.    

The spacetime given by the metric in Eq.  (\ref{sz}) is clearly not asymptotically flat, but its maximal analytic extension is exactly akin to the Kruskal-Szekeres frame description of the Schwarzschild spacetime \cite{ernst1976}. The only physical singularity in the spacetime occurs at $r=0$ similar to the usual Schwarzschild spacetime. 

Curiously, the Kretschmann scalar $K \equiv R^{\mu \nu \rho \lambda} R_{\mu \nu \rho \lambda}$ corresponding to the spacetime of Eq.  (\ref{sz}) has the property that $\lim_{r \rightarrow \infty, \theta \rightarrow \pi/2} K = 0$ (see Appendix \ref{kms}). To reinforce this conclusion, a verification check has been performed on the Newman-Penrose curvature scalar coefficients. For $\theta=\pi/2, r \rightarrow\infty$, we get $\Phi_{01}= \Phi_{02} = \Phi_{12}= 0 = \Phi_{00}=\Phi_{11} =\Phi_{22} =  \Lambda_{NP} = \Psi_0 = \Psi_1 = \Psi_2 = \Psi_3 = \Psi_4$ \footnote{We put a subscript `NP' in $\Lambda_{NP}$ to show that it is completely different from $\L$ of Eq.  (\ref{oL}).}. Likewise the electromagnetic field invariant ($F_{\mu\nu}F^{\mu\nu}$) also vanishes asymptotically in the same limit on the equatorial plane, as can be verified from the expression derived in \cite{gp1978}. These are characteristics of a restricted domain of asymptotic flatness in an otherwise asymptotically non-flat geometry. It also reinforces the idea that the precession effects are confined to mostly the emission region where a non-trivial magnetic field is discernible.  

\subsection{Fundamental precession frequencies\label{s3}}

The three fundamental frequencies which are related to the orbit of a test particle, are very important for the accretion disk theory, and are directly derived from the metric components. These are, the Keplerian frequency $\O_{\phi}$, vertical epicyclic frequency $\O_{\th}$, and the radial epicyclic frequency $\O_r$. The latter two frequencies are related  to the precession of the orbit and orbital plane. Precession of the orbit is  measured by the periastron precession frequency $(\O_{\rm per})$, and the nodal plane precession is measured by the angular frequency $\O_{\rm nod}$. To derive the epicyclic frequencies, one can consider the small perturbations \cite{don, don2} 
\begin{eqnarray}
 r(t)=r+\delta r(t)
 \\
 \th(t)= \f{\pi}{2}+\delta \th (t)
\end{eqnarray}
of a stable circular orbit of radius $r$ in the equatorial plane $\th=\pi/2$. General expressions for the three fundamental frequencies for the stationary and axisymmetric spacetime, in terms of the metric components, were obtained by Ryan\cite{ryan}. In case of a static spacetime ($g_{t\phi}=0$), one can simplify the expressions to 
\begin{eqnarray}
\O_{\phi} &=& \sqrt{-\f{\p_r g_{tt}}{\p_r g_{\phi\phi}}}\Bigg|_{\th=\pi/2},
\label{otp}
\\
\O_r &=& \f{1}{\sqrt{2g_{rr}}}\left[g_{tt}^2\p_{r}^2(1/g_{tt})+\O_{\phi}^2 g_{\phi\phi}^2 \p_{r}^2(1/g_{\phi\phi}) \right]^{\f{1}{2}}\bigg|_{\th=\pi/2}
\label{otr}
\end{eqnarray}
and
\begin{eqnarray}
\O_{\th} &=& \f{1}{\sqrt{2g_{\th\th}}}\left[g_{tt}^2\p_{\th}^2(1/g_{tt})+\O_{\phi}^2 g_{\phi\phi}^2 \p_{\th}^2(1/g_{\phi\phi}) \right]^{\f{1}{2}}\bigg|_{\th=\pi/2}.
\label{ott}
\end{eqnarray}
The other two important frequencies are defined as \cite{bs}
\begin{eqnarray} 
\O_{\rm per}= \O_{\phi}-\O_r, 
\label{per}
\\
\O_{\rm nod}= \O_{\phi}-\O_{\th}.
 \label{no} 
\end{eqnarray}
It was earlier shown by many authors that the expressions of $\O_{\phi}$ and $\O_{\th}$ match each other in any non-rotating spacetime, and, hence, $\O_{\rm nod}$ vanishes. For example, it vanishes in the regular Schwarzschild spacetime, whereas, it does not vanish in the Kerr spacetime. The so-called LT precession frequency (or $\O_{\rm nod}$) was derived \cite{lt} by Lense and Thirring for the slowly rotating Kerr spacetime. Later, Kato \cite{ka} derived the exact expression for $\O_{\rm nod}$ in the Kerr spacetime. It is clear that the nodal plane precession is {\it not} vanishingly small for the EMS spacetime. We now turn to the actual computation of these angular frequencies.

\section{Angular frequencies for the EMS spacetime \label{s3}}

\subsection{Analytic results for angular frequencies}

Let us now derive the three fundamental frequencies for the magnetized Schwarzschild spacetime. Using Eqs.  (\ref{otp}--\ref{ott}) one can obtain the Kepler frequency as ($w = r_s/r_B~,~x = r/r_B$)
\begin{eqnarray}
 \O_{\phi} &=& \f{1}{\sqrt{2}~r_s}\left(\f{w}{x} \right)^{\f{3}{2}}\left[(1+x^2)^2 \left(\f{1-3x^2+\f{4x^3}{w}}{1-x^2} \right)^{\f{1}{2}} \right] \nonumber \\
 &=& \f{(4+B^2 r^2)^2}{16} \sqrt{\f{M(4-3B^2r^2)+2B^2r^3}{r^3(4-B^2r^2)}}
 \label{ope1}
\\
&\approx& \sqrt{\f{M}{r^3}}\left[1+\f{1}{4}B^2r^2 \left(1+\f{r}{M} \right)\right] + \mathcal{O}(B^3)  \nonumber
\end{eqnarray}
and, the radial epicyclic frequency as
\begin{eqnarray}
 \O_r &=& \f{1}{r^2 (4 + B^2 r^2)}\sqrt{\f{N}{(4 - B^2 r^2)}} ,
 \label{ore1}
\end{eqnarray}
where 
\begin{eqnarray}\nonumber
 N &=& 4 B^2 r^4 (32 - 12 B^2 r^2 + 3 B^4 r^4) \nonumber
 \\
 &-& M^2 (384 - 672 B^2 r^2 + 200 B^4 r^4 - 30 B^6 r^6) \nonumber
 \\
 &+& M r (64 - 624 B^2 r^2 + 204 B^4 r^4 - 37 B^6 r^6). \nonumber
 \\
\end{eqnarray}
The vertical epicyclic frequency is obtained as
\begin{eqnarray}
  \O_{\th} &=& \sqrt{\f{M}{r^3}} \label{omth}
\end{eqnarray}

Eqs. (\ref{ope1}) and (\ref{ore1}) exhibit a pathology at $r=r_B$ : the angular frequencies diverge at this value of the radial coordinate and all turn {\it imaginary} for $r > r_B$. It follows that $r=r_B$ signifies a threshold beyond which each precession mode exhibits a quasi-normal mode-like behaviour : the precession decays to zero for sufficiently large values of $r$. Thus the precession can be physically relevant only in the emission region $r \in [r_s,r_B]$.  

Even though we have designated the emission region generically in the range $[r_s,r_B]$, strictly speaking though observations extend inward towards $R_h$ only so far as the Innermost Stable Circular Orbit radius $R_{\rm ISCO}$. In other words, for all practical purposes, the emission region can be taken to be $r \in [R_{\rm ISCO}, r_B]$.  
 
\subsection{Innermost stable circular orbit in the EMS black hole\label{s3.1}}

To determine $R_{\rm ISCO}$, we follow \cite{don}, and set $\O_r^2=0$ since for circular orbits there is no radial motion of test partcles. This implies that $R_{\rm ISCO} \equiv r_I$ must satisfy
\begin{eqnarray} \nonumber
&&12 B^6 r_I^8- 37 B^6 M r_I^7+ (30 B^6 M^2-48 B^4) r_I^6  \nonumber
\\
&+& 204 B^4 M r_I^5+(128 B^2 - 200 B^4 M^2) r_I^4 -624 B^2 M r_I^3\nonumber
\\
 &+&  672 B^2 M^2 r_I^2+ 64 M (r_I-6M)=0,
 \label{isco}
\end{eqnarray}
such that the solution of this equation would yield the ISCO radius $r_I=r_I(B,M)$. If we change to dimensionless variables $x_I \equiv r_I/r_B$, it is then straightforward to verify that Eq.  (\ref{isco}) can be reexpressed as 
\begin{eqnarray}
\sum_{n=0}^8 a_n(w) x_I^n = 0 ~\label{sisco}
\end{eqnarray}
where the coefficients of the eighth order polynomial equation $a_n=a_n(w)~,~n=0,1,...8$ for $w=r_s/r_B = BM$. This of course implies that the dimensionless ISCO radius $x_I=x_I(w) =x_I(BM)$, i.e., it is a function of the single variable $BM$ rather than being a function separately of $B$ and $M$. It is obvious that the Schwarzschild limit ($B=0$) yields $r_I=3r_s$, as expected. However, it is clear that Eq.  (\ref{sisco}) has no analytic solution, and must be solved numerically. The actual functional form of the $\{ a_n(w) \}$ are given by
\begin{eqnarray}
a_0 = -3 w^2~,~ a_1 &=& w~,~a_2 = 21 w^2~,~ \nonumber \\
a_3 = -39w~,~a_4 = 16 &-& 25 w^2~,~a_5 = 51 w~,~ \nonumber \\ 
a_6=15 w^2-24,~ a_7 &=&-37w,~a_8=24.    
\end{eqnarray}   
We can solve $x_I$ as a function of the dimensionless variable $w$ to get an idea as to its behaviour over the range of astrophysical interest.
\begin{figure}[h!]
 \begin{center}
 {\includegraphics[width=2.8in,angle=0]{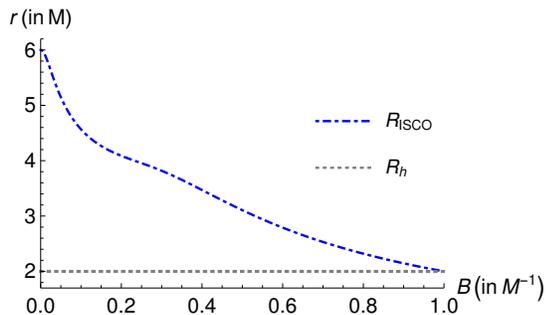}}
\caption{\label{f1}Plot of the ISCO radius $R_{\rm ISCO}$ (in $M$) for the various values of $B$ (in `$M^{-1}$'). Although $R_{\rm ISCO}$ coincides with horizon at $R_h$ for $B=M^{-1}$, i.e., for $w=1$, it disappears behind $R_h$ for $B > M^{-1}$. See section \ref{s3.1} for details.}
\end{center}
\end{figure}

The ISCO radius for a regular Schwarzschild BH (with $B=0$) is generally obtained as $6M$ in the geometrized unit. One can also obtain the same value, $R_{\rm ISCO}=6M$ from Eq.  (\ref{isco}) for $B=0$, whereas for $B=M^{-1}$ the ISCO coincides with the horizon:$R_h=2M$. For $B > M^{-1}$, the ISCO occurs at $r < 2M$, in principle. Thus, all orbits at $r > 2M$ are stable in that sense. The dot-dashed blue curve in figure \ref{f1} shows how the ISCO radius varies with the magnetic field parameter $B$. If the value of $B$ increases from $0$ to $M^{-1}$ and more, the ISCO radius decreases from $6M$ to $2M$ and then disappears behind the horizon. For weak magnetic field, i.e., $B << M^{-1}$, the ISCO occurs close to $6M$ but the exact value is lower than $6M$ \cite{shy}.

Now, in this respect, we recall that the parameter $r_B$ in our spacetime metric is roughly the extent of the region over which the magnetic field a uniform. For a weakly accreting SMBH like Sgr A*, this is identified with the region over which the radio pulsar emission undergoes Faraday rotation. Likewise for an active galactic nucleus like M87*, we may identify this as the extent of the region of emission of synchrotron radiation which also undergoes Faraday rotation. Now, according to \cite{eatough2013, eht2021}, this region is approximately $5-10$ times the Schwarzschild radius. Adding more leeway to this, albeit slightly ad hoc, the range of the dimensionless parameter $w$ may be chosen to be $w \in [0.01,0.1]$. For some representative values in this range, one may numerically solve for the dimensionless ISCO radius $x_I$. Referring to the footnote in Sect. \ref{s2}, we see that the observed value of the magnetic field is related to the parameter $B$ through the multiplicative factor $\lambda \in [0,1]$. It is somewhat mysterious that we need to choose $\lambda$ very small to yield observed values of the magnetic field from the parameter $B$.

\subsection{Gravitational Larmor Precession Frequency\label{glp}}

\begin{figure}[h!]
 \begin{center}
 {\includegraphics[width=2.8in,angle=0]{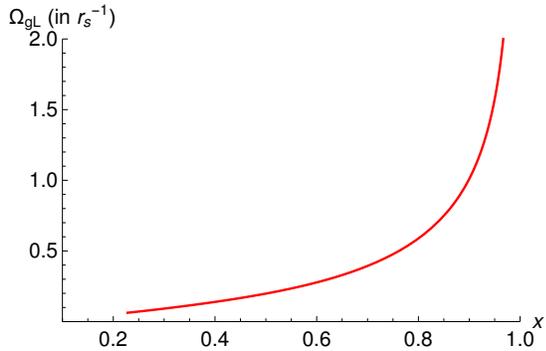}}
\caption{\label{f2}$\O_{\rm gL}$ (in $r_s^{-1}$) vs $x$ for $w=0.1$ (i.e., $B=0.1M^{-1}$) with $x_I=0.23$ (i.e., $R_{\rm ISCO}=4.57M$). See Sect. \ref{glp} for details.}
\end{center}
\end{figure}

Our interest, as already mentioned, is in the frequency of precession of the nodal plane $\O_{\rm nod}$ defined in Sect. \ref{s3}, which we call as the gravitatiotal Larmor precession ($\O_{\rm gL}$) frequency. In terms of our dimensionless variables, this is given by 
\
\begin{eqnarray}\nonumber
& \Omega_{\rm nod} & \equiv   \O_{\rm gL}
= \f{1}{\sqrt{2} r_s}\left(\f{w}{x} \right)^{\f{3}{2}} \cdot
\\
&& \Big[ (1+x^2)^2 \left(\f{1-3x^2+\f{4x^3}{w}}{1-x^2} \right)^{\f{1}{2}} 
- 1 \Big].
\label{new}
\end{eqnarray}
As the EMS spacetime is axisymmetric, the precession axis of the orbital/nodal plane should be considered along the direction of the magnetic field, i.e., the perpendicular to the orbital plane.
Observe that $\Omega_{\rm gL}$ scales with the Schwazschild radius, so that one can construct the {\it dimensionless} number $\Omega_{\rm gL} r_s$ and consider its variation with the dimensionless distance $x$. To do this one may choose a value of $w$, determine the ISCO radius $x_I(w)$ for the chosen value of $w$, and consider the variation of $r_s \Omega_{\rm gL}$ as a function of $x \in [w,1)$. For example, we show the variation of $\O_{\rm gL}$ vs $x$ in figure \ref{f2} for $w=0.1$ with $x_I=0.23$. The maximum value of the domain interval is not considered because that corresponds to $\O_{\rm gL}$ blowing up before turning imaginary. However, the intriguing feature is that the GLP frequency increases (i.e., approximately, $\O_{\rm gL} \propto x^{3/2}$) with the  increasing of $x$ from $r_I$ to $R_B$. This is not so unusual, as we have  already seen such a similar trend of the precession frequency (i.e., $\O_{\rm nod}$ increases with increasing of $r$ in some specific ranges of $r$) in case of the Kerr naked singularity (see figures. 8 and 9 of \cite{ckp}). It is also exciting that the magnetic field is solely responsible for the generation of GLP, as Eq.  (\ref{new}) shows that the GLP can arise even if the mass of the spacetime tends to zero.

Note that the value of $\O_{\rm gL}$ could be extremely small for a typical BH, as its surrounding magnetic field is found to be extremely small compared to $B_M \sim 10^{19}M_{\odot}/M$ Gauss (see Eq.  1.2 of \cite{gp1978}). However, the value of $\O_{\rm gL}$ could be much higher in case of a magnetar and could be measurable from the astrophysical observation. As the spin of a magnetar is found to be very small ($\sim 1$ Hz), the LT precession rate ($\O_{\rm LT}$) should be negligible. On the other hand, $\O_{\rm gL}$ could be comparatively much higher than $\O_{\rm LT}$, as the magnetic field is around $10^{15}$ Gauss close\footnote{One may consider an approximately uniform magnetic field in this range, where $\O_{\rm gL}$ could be measured by a(n) (in)direct astrophysical observation. Note that this is just a crude estimate to give an idea for measuring the GLP. For an exact calculation of $\O_{\rm gL}$ around a magnetar, one has to start from a proper metric describing the magnetar, since Eq.  (\ref{sz}) is not exactly applicable for a magnetar.}($r \sim 10 - 10^4$ km) to a typical magnetar. Therefore, if it is possible to measure the orbital plane precession rate indirectly (i.e., by measuring the QPO frequency \cite{sv, bs} and so on) for a magnetar, it might have arisen due to $\O_{\rm gL}$, not for the so-called $\O_{\rm LT}$ which arises due to the spin of the magnetar.

\section{Discussion\label{dis}}

We have shown that the gravitational Larmor precession, which is similar to the LT effect in the Kerr spacetime, arises in the Schwarzschild spacetime if it is immersed in a magnetic field. Unlike \cite{agmp}, no charged particles motion is involved in our analysis. Similarly, no test spinning particles motion is also considered here unlike \cite{ram, deri}. Only magnetic field is responsible for this precession. If the magnetic field vanishes (i.e., $w \rightarrow 0$), the GLP becomes zero. One can generalize the formulation presented in this paper for a slowly rotating Kerr BH with a magnetic field in its vicinity, i.e., the GLP can also arise in the magnetized Kerr spacetime. As the LT effect ($\O_{\rm nod}$) is already present in the Kerr spacetime, the sole effect of GLP has to be separated from the LT precession. The nodal plane precession discussed by us arises only due to the presence of a proximitous magnetic field, when no Kerr parameter is present. The generalization to the case of a rotating Kerr spacetime in presence of a magnetic field, is under preparation, and will be discussed elsewhere, hopefully in the near future. 

We should point out however a basic difference between the GLP considered here, and the LT precession. The LT effect is not only responsible for the orbital plane precession \cite{lt, ckp, cb17, bcb19}, but it also responsible for the precession of test gyroscopes \cite{sf, cm, ckj, cc} and the `gravitational Faraday effect' \cite{nouri, cprd1} in Kerr spacetime. On the other hand, the GLP occurs only for the orbital plane precession in the presence of magnetic field in a {\it non-rotating} spacetime, without the concomitant occurrence of LT precession of test gyroscopes. This is the hallmark of the phenomenon analyzed here. 
We may compare our findings with similar effects involving magnetars, neutron stars with prominent magnetic fields, some of which present themselves as radio or X-ray pulsars. 

The metric dealt with in this paper is not asymptotically flat, as mentioned earlier. This is, in fact, a standard practice to
consider a uniform magnetic field (in the beginning) around a BH to explain the several astrophysical phenomena (e.g. see \cite{cprd1, shi, za, za2, tur, nia, dad} and so on), as it is easier to handle. However, as the measuring of C-type low frequency QPO (LF QPO) frequency (see \cite{bs, sv, cbgm} and Sect. VII.A of \cite{ckp} for more details about the method) is sufficient to obtain the orbital plane precession rate \cite{bs, sv}, one does not need any separate and special treatment to measure the GLP frequency. This is because of the fact that the GLP is also an orbital/nodal plane precession. The implication of GLP in the field of Astrophysics can be huge. For instance, Bardeen-Petterson (BP) effect \cite{bp} plays an important role in the accretion disk theory, which arises due to the orbital plane precession, or, the so-called LT effect that emerges due to the non-zero Kerr parameter \cite{cb17}. In a stark contrast, our result shows that one can obtain a non-zero orbital plane precession in the mere presence of a magnetic field. We emphasize here that the presence of Kerr parameter is not one and only criteria to obtain a non-zero orbital plane precession as well as the BP effect. The BP effect can arise in case of a non-rotating BH if it is surrounded by a non-zero magnetic field. We wish to take that up in detail elsewhere.

\onecolumngrid

\begin{appendix}

\section{\label{kms}Kretschmann scalar} 
The Kretschmann scalar ($K$) for the EMS spacetime is calculated as:
\begin{eqnarray}\nonumber
 K &=& \f{4}{r^6 r_B^4 \L^8}\left[224 (r^3 + r_B^2 r_s)^2 + 
 16 (-12 r^6 - 46 r^3 r_B^2 r_s + 9 r r_B^4 r_s - 43 r_B^4 r_s^2) \L \right.
 \\ \nonumber
&+& \left.  4 (12 r^6 + 96 r^3 r_B^2 r_s - 72 r r_B^4 r_s + 191 r_B^4 r_s^2)\L^2-36r_s r_B^2 (2 r^3 - 5 r r_B^2 + 10 r_B^2 r_s) \L^3 \right.
\\ 
&-& \left. 9 (4 r - 7 r_s) r_s r_B^4 \L^4 \right] .
 \label{kret}
\end{eqnarray}
 Eq. (\ref{kret}) vanishes in the range of $0 < \th \leq \pi/2$ at $r \rightarrow \infty$. At $(\th = 0, r \rightarrow \infty)$, it reduces to 
\begin{eqnarray}
 K \Big|_{(\th = 0, r \rightarrow \infty)} = \f{320}{r_B^4}
\end{eqnarray}
which is finite. Eq. (\ref{kret}) reduces to 
\begin{eqnarray}
 K_{\rm Schwarzschild}=\f{12r_s^2}{r^6}
\end{eqnarray}
for $B \rightarrow 0$, or, $r_B \rightarrow \infty$.

\section{\label{nps}Newman-Penrose curvature scalar coefficients}
In the Newman-Penrose (NP) formalism, independent components of the Ricci tensors are encoded into seven (or ten) Ricci scalars which consist of three real scalars ($\Phi_{00},\Phi_{11},\Phi_{22}$) , three (or six) complex scalars ($\Phi_{01}=\bar{\Phi}_{10},\Phi_{02}=\bar{\Phi}_{20}, \Phi_{12}=\bar{\Phi}_{21}$) and the NP curvature scalar $\Lambda_{NP}$ \footnote{We put a subscript `NP' in $\Lambda_{NP}$ to show that it is completely different from $\L$ of Eq.  (\ref{oL}).} .  
Now, using the following adapted null tetrad
\begin{eqnarray}
 {\bf l} &=& \f{1}{\sqrt{2}}\left[\f{1}{\L\sqrt{1-r_s/r}}~\p_t+\f{\L}{r\sin\th}~\p_{\phi} \right],
 \\
 {\bf n} &=& \f{1}{\sqrt{2}}\left[\f{1}{\L\sqrt{1-r_s/r}}~\p_t-\f{\L}{r\sin\th}~\p_{\phi} \right],
 \\
 {\bf m} &=& \f{1}{\L \sqrt{2}}\left[\sqrt{1-r_s/r}~\p_r+i\f{1}{r}~\p_{\th} \right],
 \\
 {\bf{\bar{m}}} &=& \f{1}{\L \sqrt{2}}\left[\sqrt{1-r_s/r}~\p_r-i\f{1}{r}~\p_{\th} \right],
\end{eqnarray}
the above-mentioned curvature scalar coefficients for the EMS metric (Eq.  \ref{sz}) are calculated as:
\begin{eqnarray}
 \Phi_{00} &=& \Phi_{22}= \f{2(r-r_s\sin^2\th)}{r r_B^2 \L^4} ,
 \label{h00}
 \\
  \Phi_{11} &=& \Phi_{01} = \Phi_{12}=0 ,
    \\ \nonumber
     \Phi_{02} &=&  \f{-r_B^4 r_s - 2 r^2 r_B^2 r_s \sin^2\th - 
 r^4 r_s \sin^4\th - (2 r - r_s) r_B^4 \L^2 \cos 2\th + 2i\sqrt{r(r-r_s)}r_B^4 \L^2 \sin2\th}{r r_B^6 \L^6}   ,
 \\
     \\
     \L_{NP} &=& 0
     \label{lnp} .
\end{eqnarray}

The five NP Weyl scalars $(\Psi_0 , \Psi_1 , \Psi_2 , \Psi_3 , \Psi_4)$ for the EMS metric (Eq.  \ref{sz}) are calculated as:
\begin{eqnarray} \nonumber
 \Psi_0 &=& \f{6}{r^{7/2} r_B^8 \L^6} \left[i r^3 \sqrt{r-r_s}r_B^4 \L^2 (r_B^2 - r^2 \sin^2\th)\sin\th \cos\th 
 +\f{1}{2} r^{19/2} \cos 2\th \sin^6\th + \f{5}{8} r^{17/2} r_s \sin^8\th  \right.  \nonumber
 \\
 &+& \left.
 \f{1}{16} \sqrt{r}
   r_B^2 (-r_B^2 (-2 r^5 + 3 r^4 r_s + 6 r^2 r_B^2 r_s + 2 r_B^4 r_s + 
       2 r^2 (2 r^3 + 4 r r_B^2 - 2 r^2 r_s - 3 r_B^2 r_s) \cos 2\th \right.  \nonumber
       \\
       &+& \left. 
       r^4 (-2 r + r_s) \cos 4\th) + 8 r^7 \cos 2\th \sin^4\th + 
    12 r^6 r_s \sin^6\th) \right] ,
 \label{s0}
 \\
  \Psi_1 &=& \Psi_3= 0 ,
  \\
   \Psi_2 &=& \f{(r_B^2 - r^2 \sin^2\th)(4 r^3 - 3 r^2 r_s + r_B^2 r_s + 3 r^2 r_s \cos^2\th)}{4r^3 r_B^4 \L^4} ,
    \\
     \Psi_4 &=& \f{6}{r^{7/2} r_B^8 \L^6} \left[-i r^3 \sqrt{r-r_s}r_B^4 \L^2 (r_B^2 - r^2 \sin^2\th)\sin\th \cos\th 
 +\f{1}{2} r^{19/2} \cos 2\th \sin^6\th + \f{5}{8} r^{17/2} r_s \sin^8\th  \right.  \nonumber
 \\
 &+& \left.
 \f{1}{16} \sqrt{r}
   r_B^2 (-r_B^2 (-2 r^5 + 3 r^4 r_s + 6 r^2 r_B^2 r_s + 2 r_B^4 r_s + 
       2 r^2 (2 r^3 + 4 r r_B^2 - 2 r^2 r_s - 3 r_B^2 r_s) \cos 2\th \right.  \nonumber
       \\
       &+& \left. 
       r^4 (-2 r + r_s) \cos 4\th) + 8 r^7 \cos 2\th \sin^4\th + 
    12 r^6 r_s \sin^6\th) \right]
     \label{s4} .
\end{eqnarray}

It can be seen that Eqs.  (\ref{h00}--\ref{lnp}) and Eqs.  (\ref{s0}--\ref{s4}) vanish in the range of $0 < \th \leq \pi/2$ at $r \rightarrow \infty$. Thus, one can conclude that the Kretschmann scalar and the Newman-Penrose coefficients for the Ricci and Weyl scalars all vanish at the radial asymptopia $r \rightarrow \infty$ {\it on the equatorial plane} $\th \rightarrow \pi/2$. Note that, one may also calculate the Zakhary-McIntosh curvature invariants for the EMS metric following \cite{kr22}, as those allow a manifestly coordinate invariant characterisation of some geometrical properties of spacetimes, e.g. curvature
singularities, gravitomagnetism etc \cite{kr22}.
\end{appendix}



\end{document}